# *Application and practice of AI technology in quantitative investment*

**Shuochen Bi[1,a], Wenqing Bao[2,b], Jue Xiao[3,c], Jiangshan Wang[4,d], Tingting Deng[5,e]**

[1]D'Amore-McKim School of Business, Northeastern University, Independent Researcher Boston, MA, 02110, USA
[2]Americold Logistics, LLC Atlanta, GA, 30319, USA
[3]The School of Business, University of Connecticut, Independent Researcher Jersey City, NJ, 07302, USA
[4]The Paul Merage School of Business, University of California, Irvine, Independent Researcher Salt Lake City, UT, 84121, USA
[5]Simon Business School, University of Rochester, Independent Researcher Chantilly, VA, 20151, USA
[a]bi.shu@northeastern.edu, [b]bao.234@osu.edu, [c]juexiaowork@gmail.com,
[d]Jiangshanwang6@gmail.com, [e]dengtin1@gmail.com,

*Abstract:* With the continuous development of artificial intelligence technology, using machine learning technology to predict market trends may no longer be out of reach. In recent years, artificial intelligence has become a research hotspot in the academic circle, and it has been widely used in image recognition, natural language processing and other fields, and also has a huge impact on the field of quantitative investment. As an investment method to obtain stable returns through data analysis, model construction and program trading, quantitative investment is deeply loved by financial institutions and investors. At the same time, as an important application field of quantitative investment, the quantitative investment strategy based on artificial intelligence technology arises at the historic moment. How to apply artificial intelligence to quantitative investment, so as to better achieve profit and risk control, has also become the focus and difficulty of the research. From a global perspective, inflation in the US and the Federal Reserve are the concerns of investors, which to some extent affects the direction of global assets, including the Chinese stock market. This paper studies the application of AI technology, quantitative investment, and AI technology in quantitative investment, aiming to provide investors with auxiliary decision-making, reduce the difficulty of investment analysis, and help them to obtain higher returns.

## 1. Introduction

The traditional quantitative investment strategy is based on the classic financial investment theory model, uses the powerful computing power of computers to find the potential market law and value law in the financial timing data, combines the theoretical model with practical experience, and shows the advantages of quantitative investment in obtaining stable returns. Quantitative investment is based on massive financial data and effective investment theoretical model. With the help of the rapid development of data mining technology, it realizes more accurate investment decision and real-time trading, which greatly improves the accuracy and timeliness of securities investment trading.

Quantitative investment has been developing in China for more than a decade. In recent years, more and more investment institutions have announced the adoption of artificial intelligence (AI) technology. After three generations of rapid iteration and development, quantitative investment, with the support of AI technology, is currently in the new air outlet for the application of artificial intelligence technology. From a global perspective, inflation in the US and whether the Federal Reserve will stop raising interest rates are the concerns of investors, which to some extent affects the direction of global assets, including the Chinese stock market. The application of AI technology in quantitative investment strategy research has a strong application value.

## 2. The concept and development of quantitative investment

Quantitative investment refers to the trading method of issuing buying and selling orders through quantitative methods and computer programming for the purpose of obtaining stable returns.

### 2.1 The concept of quantitative investment

Quantitative investment is an investment strategy based on a large amount of historical data and mathematical models. It uses computer models, complex algorithms and other technical means to predict and analyze the market historical data to help investors quickly and efficiently develop investment strategies, aiming to reduce the investment risk and improve the return rate. Quantitative investment involves stocks, futures, foreign exchange and other fields, and the construction and adjustment of investment portfolio are based on a large number of historical data and analysis results.

Specifically, quantitative investing builds a model that can identify stocks with above average probability of performing the benchmark index. To implement a model, stocks are usually assigned a score based on one or more features (or factors) and then ranked. Quantitative portfolios typically hold top-ranked stocks and then rebalance them regularly or when inconsistent with the model.

The core of quantitative investment is to make investment decisions through scientific and objective methods, and to avoid the influence of subjective factors on investment decisions, so as to achieve a higher rate of return and a lower risk. This investment method helps investors to better understand the market trends and quotations, so as to make more accurate investment decisions, and can also avoid the interference of emotional and subjective factors.

### 2.2 The development of quantitative investment

Quantitative investment has a development history of more than 30 years overseas. Its investment performance is stable, and its market size and share are constantly expanding, which has been recognized by more and more investors. From the perspective of participants in the global market, according to the scale of assets under management, many world-renowned asset management institutions rely on computer technology to carry out investment decisions, and the scale of funds managed by quantitative and programmatic exchanges is constantly expanding. In general, quantitative investment is constantly developing and innovating in technology, data, algorithms and trading strategies, providing investors with more opportunities and challenges.

The emergence of quantitative investment can be traced back to the 1960s, when the emergence of scientific stock market systems and convertible bond arbitrage strategies marked the rise of quantitative investment. These early quantitative strategies were mainly based mainly on mathematical and statistical principles, using computer models for investment decisions. In the 1970s and 1980s, quantitative investment developed further, and option pricing theory and statistical arbitrage strategy began to rise. These theories provide a more solid theoretical basis for quantitative investment and promote the diversification of quantitative investment strategies. In the 1990s, it was called the "golden decade" of quantitative investment. The formation of standard financial theory and the prosperity of quantitative investment funds brought unprecedented development opportunities for quantitative investment. During this period, quantitative investment strategies were widely used, and many investors began to adopt quantitative methods to manage their assets.

However, in the 21st century, quantitative investment has also encountered some challenges. After 2000, the subprime mortgage crisis and other financial events had an impact on quantitative investment, the market volatility increased, and the effectiveness of quantitative strategy was questioned. But still, quantitative investment continues to evolve and innovate.

With the development of technology, especially the application of artificial intelligence and machine learning in quantitative investment, quantitative investment strategies have become more intelligent and accurate. These technologies allow quantitative models to better handle complex data, discover the investment opportunities hidden in the market, and improve the investment efficiency and return rate.

With the continuous development and improvement of the global financial market, the application scope of quantitative investment is also expanding. Nowadays, quantitative investment has been widely used in stocks, futures, foreign exchange and other markets, and has become one of the important tools for investors to manage assets.

## 3. The use of artificial intelligence in quantitative investment

### 3.1 Overview of artificial intelligence

Artificial intelligence (Artificial Intelligence, AI) is a technical science that studies and develops theories, methods, technologies and application systems for the simulation, extension and expansion of human intelligence. Some experts define AI as the science that acquiring and utilizes knowledge; others argue that AI is the intelligence work performed only by humans. Different understandings all reflect the basic ideas of artificial intelligence. They use computers to simulate human intelligent behavior, and let computers engage in the basic theories, methods and technologies of work that requires human wisdom in the past[1].

### 3.2 The development process of artificial intelligence

The concept of artificial intelligence (AI) can be traced back to the era of British mathematician and computer scientist Alan Turing (Alan Turing), when scientists tried to use machines to simulate the way humans think about handling problems to accomplish some complex tasks that only humans can perform. The emergence of computers to solve the problem of information and data storage and processing, artificial intelligence is possible to implement. John McCarthy et al in 1956. In 1957, Frank Rosenblatt invented the perceptron algorithm, as a simple neural network composed of only two layers of neurons, the perceptron was used to handle two classification tasks. Since then, more scholars have begun to devote themselves to the field of artificial intelligence.

The development process of artificial intelligence can be roughly divided into the following key

stages:

Birth stage: Before 1956, artificial neural networks and connectionism school began to breed. In 1890, The American biologist James explained for the first time the structure and function of the human brain, as well as the rules of memory, learning, association and other related functions. In 1949, Canadian psychologist Heb proposed Hebb's rules that changed the strength of neural network connections.

The first wave (preceptron): 1956 was an important year in the development of artificial intelligence. This year, the Dartmouth Conference, where scientists including McCarthy, Minsky, Shannon, Simon and Samuel gathered, marked the birth of AI.

The second wave (expert systems): Later, from 1980 to 1987, expert systems became an important direction of artificial intelligence research.

The third wave (connection mechanism): At the same time as the second wave, the connection mechanism is also a hot spot of research.

The fourth wave (deep learning): From 1993 to 2000, deep learning became a new and important direction in the development of artificial intelligence.

After entering the 21st century, with the development of network technology, especially Internet technology, artificial intelligence technology has further become practical. Now, AI has been widely used in image recognition, autonomous driving, finance, healthcare, gaming, education, home and life, logistics and supply chain management. At present, China is actively promoting the development of artificial intelligence industry, treating it as the key starting point of industrial innovation and the key engine to drive new quality productivity. It can be said that the development of artificial intelligence is entering a new stage, and its influence and application scope will continue to expand.

## 3.3 The application of AI in quantitative investment

The rapid development of artificial intelligence has a huge impact on the field of quantitative investment. As an investment method to obtain stable returns through quantitative statistics, model construction and program trading, quantitative investment is deeply loved by financial institutions and investors. As the key area of quantitative investment, the quantitative investment strategy based on artificial intelligence technology emerges. Quantitative investment based on artificial intelligence technology of financial institutions and individual investors from heavy artificial stocks analysis, entrust trading, full time see intraday liberation, the powerful analysis model, excellent investment experience, classic profit model and timely trading execution using computer implementation, avoid human error and subjective emotions, at the same time bring investors accurate analysis and automated trading experience[2].

The construction process of quantitative investment strategy based on artificial intelligence technology includes six links: data acquisition, data processing, data analysis, strategy construction, back test evaluation and strategy analysis. In addition, simulated trading and real trading can be added. Simulated trading is to further test the effectiveness of quantitative investment strategy by accessing real-time market, while real trading is to access real securities account and real-time trading to obtain profits. The construction process of quantitative investment strategy is shown in the figure 1 below:

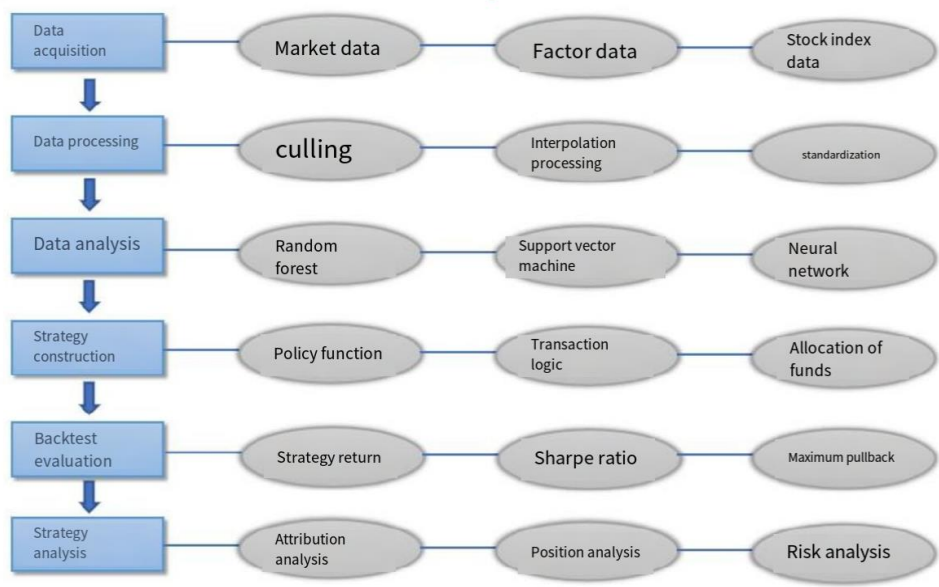

Figure 1: Quantifying the investment strategy construction process

### 3.3.1 Data acquisition

We need to obtain securities market data and fundamental data, which includes fundamental information about companies and individual stocks, factor data, stock index data, and so on. We will take factor data as the research object, and use stock index data as the benchmark for income.

### 3.3.2 Data processing

Preprocessing of data is a crucial step before conducting any data mining and analysis. By excluding delisted or suspended stocks, handling missing data, ensuring data completeness, standardizing the data, and labeling according to set rules, the quality and reliability of the data can be enhanced. Only after this preprocessing step can the data be effectively utilized for subsequent data mining and analysis activities.

### 3.3.3 Data analysis

Leveraging artificial intelligence technology for stock data analysis involves utilizing machine learning and deep learning models to mine the potential relationship between various factors and stock yields. By applying internal dynamic laws to predict future stock earnings and market performance, this approach can provide valuable data support for constructing subsequent securities portfolio strategies.

### 3.3.4 Strategy construction

To build a quantitative strategy, it is crucial to select the appropriate stock selection model and trading logic. Utilizing the data analysis process, we establish a stock selection model aimed at predicting future earnings and screening investment objects. Stocks predicted to perform well are included in the investment alternative pool, and a buying signal triggers an immediate purchase. Conversely, stocks exhibiting poor performance are sold once a sell stop signal is triggered. Additionally, the investment strategy can be customized in terms of frequency and the allocation proportion of stock capital. This approach ensures a dynamic and responsive investment strategy tailored to meet specific objectives.

### 3.3.5 Test assessment

In accordance with the construction of a portfolio investment strategy, historical market data are used to backtest specific income-generating trading strategies. Income indicators such as cumulative yield, annualized yield, excess yield, Sharpe ratio, and maximum drawdown are considered to compare the performance of different quantitative investment strategies. By evaluating the strategy's income and risk indicators, a comprehensive assessment of its effectiveness can be made.

### 3.3.6 Strategy analysis

According to the test trading results, comprehensive analysis of quantitative investment strategy, including attribution analysis, position analysis, clinch a deal analysis, risk analysis, etc., from profitability, risk control, investment style, industry selection analysis strategy, for the subsequent quantitative investment strategy improvement and optimization provide reference.

### 3.4 Artificial intelligence methods

Artificial intelligence methods are complex, mainly including random forest, support vector method, artificial neural network, attention mechanism and so on. Due to the limitation and depth of this paper, random forest is used as an example[3].

Random forest is an integrated learning algorithm containing multiple decision tree structures, whose basic unit is a single decision tree, and each decision tree is a separate classifier in the classification task. When the sample is input, each decision tree produces its own classification results. Random forest is responsible for collecting the output results of all decision trees, and takes the category with the highest voting proportion as the final result, which is the embodiment of the idea of integrated learning. If a decision tree is a weak classifier, then a random forest is a strong classifier integrated by multiple weak classifiers. The general structure of the random forest is shown in the figure 2 below:

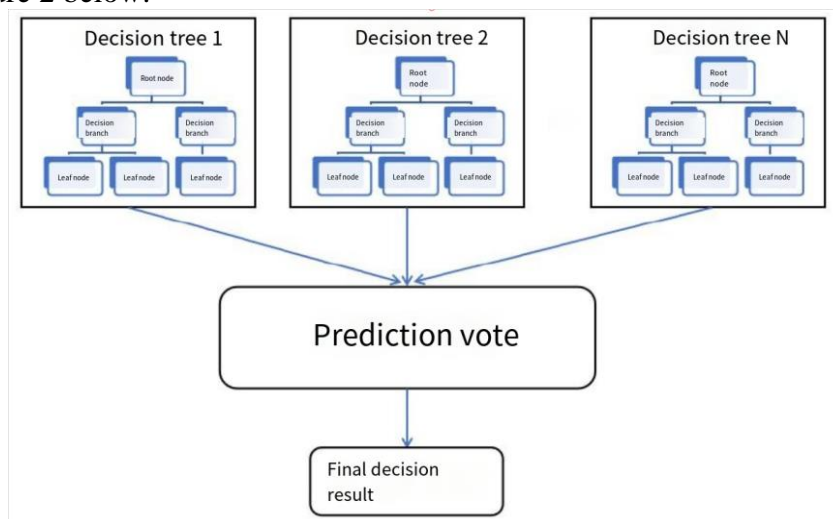

Figure 2: The approximate structure of the random forest

## 4. The role of AI technology in risk control

### 4.1 Application of AI in risk identification and assessment

AI can quickly analyze large amounts of financial data, extract key information, and judge potential risk factors. Through the iterative training of machine learning algorithms, AI can continuously optimize the risk model, identify and evaluate risks. This automated risk identification and assessment can help financial institutions detect and respond to various risks in a timely manner and reduce the losses caused by risks. In addition, AI can also monitor the financial market in real time, analyze transaction data, find abnormal transactions and give timely warning, and reduce risks such as market manipulation and insider trading. For customer risk assessment, AI can evaluate customers 'credit risk and the possibility of default through customer data analysis, help financial institutions to accurately judge customers' credit status, and reduce the risk of default and loss[4].

## 4.2 Case analysis

Case 1: Application of intelligent risk control model in stock investment

A large investment institution uses an intelligent risk control model based on AI technology to manage its equity portfolio. The model analyzes a large amount of historical data through a deep learning algorithm, identifies the key factors affecting stock prices, and constructs a prediction model. In the investment process, the model monitors the dynamic changes of the stock market in real time, and dynamically adjusts the investment portfolio to cope with the potential market risks.

At the same time, the model can also identify abnormal trading behaviors, such as abnormal stock price fluctuations, large capital flows, etc., and timely issue early warning. Investment institutions can, according to these early warning information, take corresponding risk control measures, such as adjusting positions, stop losses, etc., to avoid potential losses[5].

Through the application of the intelligent risk control model, the investment institution has achieved a good risk control effect in the stock investment, effectively reduced the investment risk, and improved the stability of the investment return.

Case 2: AI credit risk assessment in bond investment

In the field of bond investment, credit risk assessment is a crucial link. A bond investment institution uses AI technology to build a credit risk assessment model, which can accurately evaluate the credit status of the bond issuer.

The model collects multi-dimensional information such as financial data, business information and industry information of bond issuers, and uses machine learning algorithm to deeply mine and analyze these data. By constructing a complex prediction model, the model can accurately predict the default probability and credit rating of bond issuers.

In the process of investment decision-making, investment institutions can screen and allocate bonds according to the evaluation results of the model, so as to avoid investing in bonds with high credit risk. At the same time, the model can also monitor the credit risk changes in the bond market in real time, and provide timely risk early warning and coping strategies for investment institutions.

Through the application of AI credit risk assessment model, the investment institution has realized the effective risk control in the bond investment, reduced the default risk, and improved the security of the investment portfolio.

## 5. The advantages of AI technology in quantitative investment

The application of AI technology in quantitative investment can significantly improve the accuracy of prediction, optimize transaction decisions, strengthen risk management, and provide investors with real-time market monitoring and personalized investment advice. These advantages make AI technology have a wide application prospect and potential in the field of quantitative investment.

One, improved prediction accuracy: AI technologies, especially machine learning and deep learning, can process and analyze complex amounts of data and identify complex relationships between data, so as to build accurate prediction models. These models can predict market trends, price fluctuations, etc., to provide investors with their decisions.

Second, automated trading decisions: AI technology can automate trading decisions, reduce the impact of human errors and emotional decisions through algorithm trading, and improve the efficiency and speed of trading. In addition, AI can also be used to perform more complex trading strategies, such as statistical arbitrage strategies, to increase investment returns.

Third, the optimization of risk management: risk management is an important part of quantitative investment. AI technology can find out potential risk factors and predict market fluctuations through in-depth analysis of historical data, so as to formulate corresponding risk control strategies. This helps investors to better understand and manage risk and reduce investment risk[6].

Fourth, real-time market monitoring: the AI system can monitor market dynamics in real time, including price changes, trading volume, market sentiment, etc., to provide investors with real-time market information and feedback. This helps investors to quickly respond to market changes and adjust their investment strategies.

Fifth, reduce transaction costs: because AI technology can automate the execution of transaction decisions, reduce manual operations and intermediary links, thus reducing transaction costs.

Sixth, personalized investment advice: AI technology can provide customers with personalized investment advice according to investors' risk tolerance and investment objectives. This helps investors to develop the appropriate investment strategies according to their actual situation.

## 6. The challenge of applying AI technology to quantitative investment

### 6.1 Challenges and problems facing AI technology

Although the application of AI technology in quantitative investment brings many advantages, it also faces some challenges and problems, mainly manifested in:

One is data quality and source: quantitative investment is highly dependent on data, while the effect of the AI model largely depends on the accuracy and integrity of the data. Therefore, obtaining high-quality, comprehensive and reliable data is an important challenge. In addition, data privacy and security issues cannot be ignored.

The second is the model overfitting and generalization ability: the AI model may overfit in the training process, that is, the model overadapt to the training data, resulting in the decline of the predictive ability on the unknown data. At the same time, the generalization ability of the model is also a concern, that is, whether the model can maintain stable performance in different market environments and conditions.

Third, technology update and iteration: AI technology is developing rapidly, and new algorithms and models are constantly emerging. Investors need to constantly follow up on technological progress and update and optimize their investment strategies and models to respond to changes in the market.

Fourth, regulation and compliance: With the wide application of AI technology in the financial sector, regulators are also strengthening the review and supervision of AI models and strategies. Investors need to ensure that their AI models and strategies comply with relevant regulations and policy requirements to avoid the risks caused by irregularities[7].

### 6.2 Challenges facing AI technology in the context of the Fed's moratorium on interest rate hikes

In the context of the Fed's delay on interest rate hikes, the challenges of applying AI technology to quantitative investment mainly include the following aspects:

First, changes in the market environment may lead to the reduced effectiveness of the historical data. Quantitative investing is highly dependent on historical data to train and optimize AI models, but sudden changes in market conditions, such as the Fed's halt in raising interest rates, may make past data no longer fully applicable to current market conditions. This requires investors to constantly update and adjust the model to meet the new market environment.

Second, the predictive ability of the AI model may be affected by the market uncertainty. Markets could face more uncertainty and volatility after the Fed stops raising interest rates, making the AI model forecast more difficult. Investors need to be more careful about evaluating the model's predictions and making investment decisions combined with other information.

Third, investors also need to pay attention to the model risk problem. Although AI models have powerful data processing and prediction capabilities, they also have risks such as over-fitting and improper parameter adjustment. In the context of the Federal Reserve stopping raising interest rates, investors should pay more attention to the robustness and reliability of the model and avoid investment losses caused by model errors.

Finally, changes in regulatory policy may also pose challenges to the application of AI technologies to quantitative investment. With the rapid development of fintech, regulators are also strengthening the supervision of AI technology. Investors need to ensure that their AI models and strategies comply with relevant regulations and policy requirements to avoid the risks caused by irregularities.

## 7. Conclusion

The application of quantitative investment in artificial intelligence in the financial market is gradually increasing, and its accuracy and efficiency advantages have brought new opportunities and challenges to investors. With the continuous progress of technology and the increasing improvement of data, the application of artificial intelligence in quantitative investment will become more extensive and in-depth, bringing better investment returns and risk control for investors. AI technology plays an important role in the quantitative investment, and will continue to drive the development and innovation of the quantitative investment field. Investors should actively embrace AI technology, constantly learn and adapt to the changes and challenges brought about by new technologies, so as to gain advantages in the fierce market competition.

## References

*[1] Sun Shouchao. Research on the application practice of 5G + AI technology in financial media production [J]. China Broadband, 2022, 18 (5): 3.*
*[2] Wu Jing, Zhang Sen, Liu Hongbo. Application and countermeasures of blockchain technology in bidding and bidding [J]. Engineering and Management Science, 2022, 4 (2): 159-161.*
*[3] Zhao Hongye. Research on the application of AI technology in computer network technology [J]. Information Industry Report, 2023 (9): 0138-0140.*
*[4] Liu Yanhui. Analysis of the application of big data and AI technology in new media communication channels [J]. China Media Technology, 2022 (5): 70-72.*
*[5] Zhang Huiguo. AI and machine learning in financial decision-making and risk management [J]. Today, 2023 (20): 0090-0093.*
*[6] Zhu, M., Zhang, Y., Gong, Y., Xu, C., & Xiang, Y. Enhancing Credit Card Fraud Detection: A Neural Network and SMOTE Integrated Approach.[J]Journal of Theory and Practice of Engineering Science,(2024).4(02), 23–30. https://doi.org/10.53469/jtpes.2024.04(02).04*
*[7] Mengran Zhu, Ye Zhang, Yulu Gong, Kaijuan Xing, Xu Yan, Jintong Song. Ensemble Methodology: Innovations in Credit Default Prediction Using LightGBM, XGBoost, and LocalEnsemble[J]Computational Engineering, Finance, and SciencearXiv:2402.17979v1 [cs.CE] 28 Feb 2024 https://doi.org/10.48550/arXiv.2402.17979*